\documentclass[11pt, letterpaper]{article}

\usepackage{graphicx}
\usepackage{amsmath}
\usepackage{amssymb}
\usepackage{setspace}
\usepackage[letterpaper,left=1in,right=1in,top=1in,bottom=1in]{geometry}
\usepackage{color,soul}

\usepackage{hyperref}

\usepackage{subfigure}
\usepackage{float}
\usepackage{accents}

\newtheorem{remark}{Remark}

\title{\Large \bf Simultaneous state and parameter estimation: the role of sensitivity analysis}

\author{
    \centerline{\normalsize Jianbang Liu$^{a,b,c,d}$, Aristarchus Gnanasekar$^{d}$, Yi Zhang$^{e}$, Song Bo$^{d}$, Jinfeng Liu$^{d,}$\thanks{Corresponding author: J. Liu. Tel: +1-780-492-1317. Fax: +1-780-492-2881. Email: jinfeng@ualberta.ca.},}\\
    \centerline{\normalsize Jingtao Hu$^{a}$, Tao Zou$^{f}$
}\vspace{5mm}\\
\centerline{\small $^{a}$Key Laboratory of Networked Control Systems, Shenyang Institute of Automation, }\\
\centerline{\small Chinese Academy of Sciences, Shenyang 110016, China}\\
\centerline{\small $^{b}$Institute for Robotics \& Intelligent Manufacturing, Chinese Academy of Sciences,}\\
\centerline{\small Shenyang 110016, China}\\
\centerline{\small $^{c}$University of Chinese Academy of Sciences, Beijing 100049, China}\\
\centerline{\small $^{d}$Department of Chemical \& Materials Engineering, University of Alberta, Edmonton, }\\
\centerline{\small AB T6G 1H9, Canada}\\
\centerline{\small $^{e}$Key Laboratory of Energy Thermal Conversion \& Control, Southeast University,}\\
\centerline{\small Nanjing 210096, China}\\
\centerline{\small $^{f}$School of Mechanical and Electrical Engineering, Guangzhou University,}\\
\centerline{\small Guangzhou 510006, China}
}

\allowdisplaybreaks

\begin{document}

\date{}
\maketitle
\setstretch{1.35}

\begin{abstract}
State and parameter estimation is essential for process monitoring and control. Observability plays an important role in both state and parameter estimation. In simultaneous state and parameter estimation, the parameters are often augmented as extra states of the original system. When the augmented system is observable, various existing state estimation approaches may be used to estimate the states and parameters simultaneously. However, when the augmented system is not observable, how we should proceed to maximally extract the information contained in the measured outputs is not clear. This paper concerns about simultaneous state and parameter estimation when the augmented system is not fully observable. Specifically, we first show how sensitivity analysis is related to observability of a dynamical system, and then illustrate how it may be used to select variables for simultaneous estimation. We also propose a moving horizon state estimation (MHE) design that can use the variable selection results in a natural way. Extensive simulations are carried out to show the efficiency of the proposed approach. 
\end{abstract}



\section{Introduction}\label{sec:1}

State and parameter estimation is essential for process modelling, monitoring, control, and fault diagnosis, which has been extensively applied in varies fields including petrochemical, oil refining, paper making, electric power, and aerospace \cite{aster2018parameter,lei2017classification,kravaris2013advances,primadianto2016review}.

Plenty of studies have been conducted to develop various algorithms for different applications and improved performance of state and parameter estimation \cite{primadianto2016review,yin2018state,valluru2019integrated}. Broadly speaking, these algorithms can be classified into two categories. In the first category, state estimation and parameter identification are conducted separately \cite{primadianto2016review,xu2017parameter}. In general, parameter identification is first carried out to find the model parameters and then the model with the identified parameters is used for state estimation. The parameters may also be updated every some time when new process data become available. In the second category, parameter identification and state estimation are performed at the same time. This simultaneous state and parameter estimation has attracted more and more attention due to its ability to bring better stability and estimation performance \cite{ibrir2018joint,stroud2018bayesian,haessig1997method,kamalapurkar2017simultaneous}. Particularly, simultaneous state and parameter estimation is popular in model based monitoring and control \cite{kamalapurkar2017simultaneous,valluru2019integrated,bo2020parameter}. For example, in \cite{huang2010fast}, a simultaneous estimation strategy based on moving horizon estimation (MHE) was proposed, and in \cite{rangegowda2018simultaneous}, a receding horizon Kalman filter (KF) was proposed for simultaneous parameter and state estimation. And a Bayesian estimator based on robust extended Kalman filter (EKF) and moving window was proposed to estimate state and unmeasured parameter variations \cite{valluru2018development}. Among them, augmenting the parameters as extra states is a relatively common approach in simultaneous state and parameter estimation \cite{wang2009state,li2011estimation,bo2020decentralized}. For example, in \cite{li2011estimation}, ensemble Kalman filtering (EnKF) was applied to an augmented system for simultaneously estimate the calibrated soil hydraulic states and parameters. In \cite{bo2020parameter}, the application and comparison of EKF, EnKF, and MHE in simultaneous estimation were studied based on an augmented 1D infiltration process. In this work, we focus on simultaneous state and parameter estimation based on augmenting the parameters as extra states.

In parameter identification and state estimation, observability plays an important role. It is, however, challenging to check the observability of a nonlinear system directly, which involves the calculation of high-order Lie derivatives and could be very computationally demanding and is sensitive to noise \cite{marino1995nonlinear,zhang2013lyapunov,villaverde2019observability}. In applications, the observability of a nonlinear system is typically examined using approximations including linearization of the nonlinear system \cite{jayasankar2009identifiability,zeng2016distributed,nahar2019parameter} and sensitivity analysis of the nonlinear system \cite{yao2003modeling,kravaris2013advances,fysikopoulos2019framework}. In addition to the above methods, the structure of a system may also be used to examine its observability \cite{jayasankar2009identifiability,stigter2017observability}. However, structural observability may not give the degree of observability of a system. In \cite{grubben2018controllability}, the controllability and observability of 2D thermal flow in bulk storage facilities was discussed and the relation between sensitivity and observability was explored for the process. In \cite{fysikopoulos2019framework}, sensitivity was used to investigate the estimability of crystallizaton processes. In \cite{yao2003modeling}, the role of sensitivity in estimability analysis and parameter selection was discussed and applied to a gas-phase ethylene copolymerization process. While sensitivity analysis has been found to be useful and effective either in control or parameter estimation applications in the above studies, its role in simultaneous state and parameter estimation has not been studied in a systematical way.

In this work, we aim to explore the role of sensitivity analysis in simultaneous state and parameter estimation when the parameters are augmented as extra states. In particular, we are interested in the case when the augmented system is not fully observable and try to explore how we may use sensitivity analysis to pick the appropriate variables to estimate so that we can maximally extract the information contained in the measured outputs. Specifically, we will first show how sensitivity analysis is related to observability of a dynamical system, and then illustrate how sensitivity analysis may be used to select variables for simultaneous estimation. We also propose a MHE design that can use the variable selection results in a natural way. Extensive simulations will be used to show the effectiveness of the proposed approach. 


\section{Preliminaries}\label{sec:2}

\subsection{System description}\label{sec:2.1}

In the paper, we consider a general class of discrete-time nonlinear systems described as follows:
\begin{subequations}\label{eq:2.1.1}
\begin{align}
x(k+1) &= F(x(k),u(k),\theta) \label{eq:2.1.1a}\\
y(k) &= H(x(k),\theta) \label{eq:2.1.1b}
\end{align}
\end{subequations}
where $x(k) \in R^n$, $u(k) \in R^m$, and $y(k) \in R^r$ denote the state, input, and output of the system at time $k$, respectively; $\theta \in R^p$ is the parameter vector; $F(\cdot)$ and $H(\cdot)$ denote the nonlinear state and output equations, respectively. It is considered that the parameter vector $\theta$ is constant and does not change over time.

In the paper, we consider how to select states and parameters in simultaneous state and parameter estimation when not all the elements of the state and parameter vectors can be estimated simultaneously. Specially, we focus on exploring the role of sensitivity analysis in determining the most estimable state and parameter subset based on given output measurements and how it may be used to improve online state and parameter estimation performance in the framework of moving horizon estimation.

\subsection{Sensitivity evaluation}\label{sec:2.2}

In the subsection, we discuss how to calculate the sensitivity of the output $y(k)$ with respect to the initial state $x(0)$, $S_{y,x(0)}(k) := \frac{\partial y(k)}{\partial x(0)}$, and the sensitivity of the output $y(k)$ with respect to the parameter $\theta$, $S_{y,\theta}(k) := \frac{\partial y(k)}{\partial \theta}$. We will show and explore the role of the sensitivities in state and parameter estimation in the remainder of this work.

\subsubsection{Indirect approach}\label{sec:2.2.1}

In some cases, it is possible to obtain the analytical solution of system (\ref{eq:2.1.1}) and express $y(k)$ in terms of the initial state $x(0)$ and the parameter $\theta$ explicitly. If such an expression of the output $y(k)$ can be obtained, then the two sensitivities $S_{y,x(0)}(k)$ and $S_{y,\theta}(k)$ can be evaluated in a straightforward manner. Let us denote such an expression of $y(k)$ as follows:
\begin{equation}\label{eq:3.1.2}
\begin{array}{l}
y(k) = {H_{k,0}}(x(0),u(0:k-1),\theta )
\end{array}
\end{equation}
where $u(0:k-1)$ denotes the given input sequence from instant 0 to $k-1$. Then $S_{y,\theta}(k)$ and $S_{y,x(0)}(k)$ can be evaluated as below:
\begin{subequations}\label{eq:3.1.3}
  \begin{align}
    S_{y,\theta}(k) &= {\left. {\frac{{\partial {H_{k,0}}}}{{\partial \theta }}} \right|_{x(0),u(0:k-1),\theta }} \label{eq:3.1.3a}\\
    S_{y,x(0)}(k) &= {\left. {\frac{{\partial {H_{k,0}}}}{{\partial x(0)}}} \right|_{x(0),u(0:k - 1),\theta }} \label{eq:3.1.3b}
  \end{align}
\end{subequations}

However, the above explicit expression of the output shown in Eq. (\ref{eq:3.1.2}) is in general challenging to obtain. A more feasible approach is to numerically approximate the sensitivities locally about given initial state, input trajectory, and nominal parameter values by performing many experiments. In the experiments, the elements of $x(0)$ and $\theta$ are perturbed one at a time to obtain the output of the system and the sensitivities can be approximated, for example, using forward finite difference approximation as shown below:
\begin{subequations}\label{eq:3.1.4}
  \begin{align}
    S_{y,\theta_j}(k) &\approx \frac{y(k)|_{\theta_j+\Delta \theta_j}-y(k)|_{\theta_j}}{\Delta \theta_j} \label{eq:3.1.4a} \\
    S_{y,x_i(0)}(k) &\approx \frac{y(k)|_{x_i(0)+\Delta x_i(0)}-y(k)|_{x_i(0)}}{\Delta x_i(0)} \label{eq:3.1.4b}
  \end{align}
\end{subequations}
where $\theta_j$, $j = 1,2,\cdots,p$, denotes an element in the parameter vector $\theta$ and $x_i(0)$, $i = 1,2,\cdots,n$, denotes an element of the initial state vector $x(0)$; $y(k)|_{\theta_j}$ and $y(k)|_{x_i(0)}$ denote the output values at time $k$ with nominal $\theta_j$ and $x_i(0)$, respectively; $y(k)|_{\theta_j+\Delta \theta_j}$ and $y(k)|_{x_i(0)+\Delta x_i(0)}$ are the output values at time $k$ with parameter $\theta_j$ perturbed by $\Delta \theta_j$ and $x_i(0)$ perturbed by $\Delta x_i(0)$, respectively. The drawback of the experimental approximation is that a large number of experiments may be needed when the number of elements in the initial state and the parameter vectors is big.

\subsubsection{Direct approach}\label{sec:2.2.2}

The two sensitivities can also be evaluated directly without obtaining an explicit expression of the output $y(k)$ in terms of $x(0)$ and $\theta$ \cite{fysikopoulos2019framework,grubben2018controllability}. First, let us focus on ${S_{y,\theta}}(k)$ and define the sensitivity of the state to the parameter as follows:
\begin{equation}\label{eq:3.2.1}
    {S_{x,\theta}}(k) := \frac{{\partial x(k)}}{{\partial \theta }}
\end{equation}

Based on Eqs. (\ref{eq:2.1.1}) and (\ref{eq:3.2.1}), we can obtain the following two matrix equations:
\begin{subequations}\label{eq:3.2.2}
  \begin{align}
    {S_{x,\theta}}(k + 1) &= {{\frac{{\partial F}}{{\partial x}}}(k)}{S_{x,\theta}}(k) + {{\frac{{\partial F}}{{\partial \theta }}}(k)} \label{eq:3.2.2a}\\
    {S_{y,\theta}}(k) &= {{\frac{{\partial H}}{{\partial x}}}(k)}{S_{x,\theta}}(k) + {{\frac{{\partial H}}{{\partial \theta }}}(k)} \label{eq:3.2.2b}
  \end{align}
\end{subequations}
where Eq. (\ref{eq:3.2.2a}) is a matrix finite difference equation that describes the dynamics of the sensitivity ${S_{x,\theta}}(k)$, Eq. (\ref{eq:3.2.2b}) is a matrix algebraic equation that describes the relation between ${S_{y,\theta}}(k)$ and ${S_{x,\theta}}(k)$. The sensitivity ${S_{y,\theta}}(k)$ can be obtained by solving Eq. (\ref{eq:3.2.2}) with the initial condition $S_{x,\theta}(0) = 0$.

To obtain the sensitivity of the output to the initial state $S_{y,x(0)}(k)$, we can consider the initial state as a virtual parameter of the system. Let us re-write system (\ref{eq:2.1.1}) as follows to reflect its dependence on the initial state:
\begin{subequations}\label{eq:3.2.3}
\begin{align}
x(k+1) &= F(x(k),u(k),\theta,x(0)) \label{eq:3.2.3a}\\
y(k) &= H(x(k),\theta,x(0)) \label{eq:3.2.3b}
\end{align}
\end{subequations}

Similarly, let us define the sensitivity of the state to the initial condition as below:
\begin{equation}\label{eq:3.2.4}
    {S_{x,x(0)}}(k) := \frac{{\partial x(k)}}{{\partial x(0)}}
\end{equation}
Based on Eqs. (\ref{eq:3.2.3}) and (\ref{eq:3.2.4}), we can obtain the following two matrix equations describing the dynamics of $S_{x,x(0)}(k)$ and its relation to $S_{y,x(0)}(k)$:
\begin{subequations}\label{eq:3.2.5}
  \begin{align}
    {S_{x,x(0)}}(k + 1) &= {{\frac{{\partial F}}{{\partial x}}}(k)}{S_{x,x(0)}}(k) \label{eq:3.2.5a}\\
    {S_{y,x(0)}}(k) &= {{\frac{{\partial H}}{{\partial x}}}(k)}{S_{x,x(0)}}(k) \label{eq:3.2.5b}
  \end{align}
\end{subequations}
Note that in Eq. (\ref{eq:3.2.5}), there are no terms ${{\frac{{\partial F}}{{\partial x(0) }}}(k)}$ and ${{\frac{{\partial H}}{{\partial x(0) }}}(k)}$ because $x(0)$ does not explicitly present in $F(\cdot)$ and $H(\cdot)$. The sensitivity $S_{y,x(0)}(k)$ can be obtained by solving the finite difference equation (\ref{eq:3.2.5}) with the initial condition $S_{x,x(0)}(0) = I$, where $I$ being an identity matrix with dimension $n$.


\section{Relation between sensitivity and observability}\label{sec:4}

Observability plays a critical role in state estimation. In this section, we show how sensitivity and observability are related. We will focus on linear systems first and then consider nonlinear systems.

\subsection{Linear systems}\label{sec:4.1}

Let us consider the following general discrete-time linear system:
\begin{subequations}\label{eq:4.1.1}
  \begin{align}
    x(k+1) &= A x(k) + B u(k) \label{eq:4.1.1a} \\
    y(k) &= C x(k) \label{eq:4.1.1b}
  \end{align}
\end{subequations}

System~(\ref{eq:4.1.1}) is said to be observable if the initial state of the system $x(0)$ can be determined using the inputs and outputs $u$ and $y$ from 0 to $k$. It is well known that we can check whether the system is observable by checking whether the following observability matrix is full rank or not \cite{chen1998linear}:
\begin{equation}\label{eq:4.1.2}
O = \left[ {\begin{array}{*{20}{l}}
C\\
{CA}\\
\kern 3pt \vdots \\
{C{A^{n-1}}}
\end{array}} \right]
\end{equation}
where $n$ is the size of the state vector $x$.

If the above observability matrix is full rank, the system is observable and we can uniquely determine the initial state based on the input and output data. This also implies that we can estimate current state $x(k)$ of the system based on the input and output data. If the matrix is not full rank, it implies that the system is not observable and the current state $x(k)$ can not be fully estimated. 


Let us also evaluate the output to initial state sensitivity $S_{y,x(0)}$ for system (\ref{eq:4.1.1}). Using Eq. (\ref{eq:3.2.5}), we can obtain the following equations:
\begin{subequations}\label{eq:4.1.3}
  \begin{align}
    {S_{x,x(0)}}(k + 1) &= A {S_{x,x(0)}}(k) \label{eq:4.1.3a}\\
    {S_{y,x(0)}}(k) &= C {S_{x,x(0)}}(k) \label{eq:4.1.3b}
  \end{align}
\end{subequations}
where $S_{x,x(0)}(0) = I$. Solving the above equation and collecting the output to initial state sensitivities from time 0 to time $n-1$, we can obtain the following sensitivity matrix $S$:
\begin{equation}\label{eq:4.1.4}
S =  \left[ {\begin{array}{*{20}{l}}
{{S_{y,x(0)}}(0)}\\
{{S_{y,x(0)}}(1)}\\
\kern 10pt \vdots \\
{{S_{y,x(0)}}(n-1)}
\end{array}} \right]
= \left[ {\begin{array}{*{20}{l}}
C\\
{CA}\\
\kern 3pt \vdots \\
{C{A^{n-1}}}
\end{array}} \right]
\end{equation}
which is the same as the observability matrix $O$ in (\ref{eq:4.1.2}). This reveals the relation between the output to initial state sensitivities and the observability matrix for linear systems.

\subsection{Nonlinear systems}\label{sec:4.2}

For nonlinear system observability test, it requires the calculation of high order Lie derivatives and their differentials, which is generally very challenging even for systems with only a few states \cite{marino1995nonlinear}. In applications, linear approximations of a nonlinear model are often used to check the local observability of the nonlinear system. One commonly used such approach is to linearize a nonlinear system successively along typical trajectories and check the observability of the linearized models \cite{busch2013state,zeng2016distributed,nahar2019parameter}. We will focus on this approximation approach and illustrate how sensitivity analysis may be used.

Let us consider that there are in total $q$ sampling points along a trajectory of system (\ref{eq:2.1.1}). For each sampling point, $k = 0, 1,\cdots,q-1$, we can linearize system (\ref{eq:2.1.1}) and obtain the corresponding linearized model:
\begin{subequations}\label{eq:4.2.1}
  \begin{align}
    {\bar x}(i+1) &= A(k) {\bar x}(i) + B(k) {\bar u}(i) \label{eq:4.2.1a} \\
    {\bar y}(i) &= C(k) {\bar x}(i) \label{eq:4.2.1b}
  \end{align}
\end{subequations}
where $A(k):=\frac{\partial F}{\partial x}(k)$, $B(k):=\frac{\partial F}{\partial u}(k)$, and $C(k):=\frac{\partial H}{\partial x}(k)$ are time-varying matrices with respect to $k$, and ${\bar x}(i) = x(i) - x(k)$, ${\bar u}(i) = u(i) - u(k)$, ${\bar y}(i) = y(i) - y(k)$. For each sampling point $k$, $k = 0, 1,\cdots,q-1$, we can get an observability matrix based on the linearized model as shown below:
\begin{equation}\label{eq:4.2.4}
    O(k) = \left[ {\begin{array}{*{20}{l}}
C(k)\\
{C(k) A(k)}\\
\kern 6pt \vdots \\
{C(k) {{A(k)}^{n-1}}}
\end{array}} \right]
\end{equation}
We can check the rank of these observability matrices along the trajectory. If all these $O(k)$, $k = 0, 1,\cdots,q-1$, are full rank, then we may conclude that the nonlinear system is locally observable along the trajectory \cite{khalil2002nonlinear}.

Next, let us consider the sensitivity along the same trajectory of system (\ref{eq:2.1.1}). Based on Eq. (\ref{eq:3.2.5}), we can write $S_{y,x(0)}$ at the sampling point $k$ based on the following equation:
\begin{equation}\label{eq:4.2.6}
    S_{y,x(0)}(k) = {{\frac{{\partial H}}{{\partial x}}} (k)} {{\frac{{\partial F}}{{\partial x}}} (k-1)} {{\frac{{\partial F}}{{\partial x}}} (k-2)} \cdots {{\frac{{\partial F}}{{\partial x}}} (0)}
\end{equation}
Based on the definitions of $C(k)$, $A(k)$ in Eq.~(\ref{eq:4.2.1}), the above Eq. (\ref{eq:4.2.6}) can be re-written as below:
\begin{equation}\label{eq:4.2.7}
    S_{y,x(0)}(k) = C(k) A(k-1) A(k-2) \cdots A(0)
\end{equation}
For each sampling time $k$, we can collect the most recent $n$ sensitivities $S_{y,x(0)}(l)$, $l=k, k-1, \ldots, k-n+1$, to form a sensitivity matrix $S(k)$ as shown below:
\begin{equation}\label{eq:4.2.8}
    S(k) =  \left[ {\begin{array}{*{20}{l}}
{{S_{y,x(k-n+1)}}(k-n+1)}\\
{{S_{y,x(k-n+1)}}(k-n+2)}\\
\kern 10pt \vdots \\
{{S_{y,x(k-n+1)}}(k)}
\end{array}} \right]
= \left[ {\begin{array}{*{20}{l}}
C(k-n+1)\\
{C(k-n+2) A(k-n+1)}\\
\kern 6pt \vdots \\
{C(k) A(k-1) A(k-2) \cdots A(k-n+1)}
\end{array}} \right]
\end{equation}

If we compare $S(k)$ in (\ref{eq:4.2.8}) and $O(k)$ in (\ref{eq:4.2.4}), we can see that $O(k)$ contains information purely from one sampling time $k$ while $S(k)$ contains similar composition of information but from $n$ consecutive sampling times from $k-n+1$ to $k$. It is nature to expect that the rank of $S(k)$ can also be used as an indicator of the local observability of nonlinear systems. This is indeed the case. $S(k)$ has been used in many studies as an indication of the observability of nonlinear systems especially in studies on parameter selection \cite{brun2001practical,stigter2017observability,joubert2020efficient}. Note that in $S(k)$ in (\ref{eq:4.2.8}), the information from $n$ sampling times is included. It is possible to include information from more sampling points as illustrated in the simulations in Section~\ref{sec:7}.

\section{Simultaneous estimation integrated with sensitivity analysis}\label{sec:6}

\begin{figure}
  \centering
  \includegraphics[width=0.75\textwidth,angle=0]{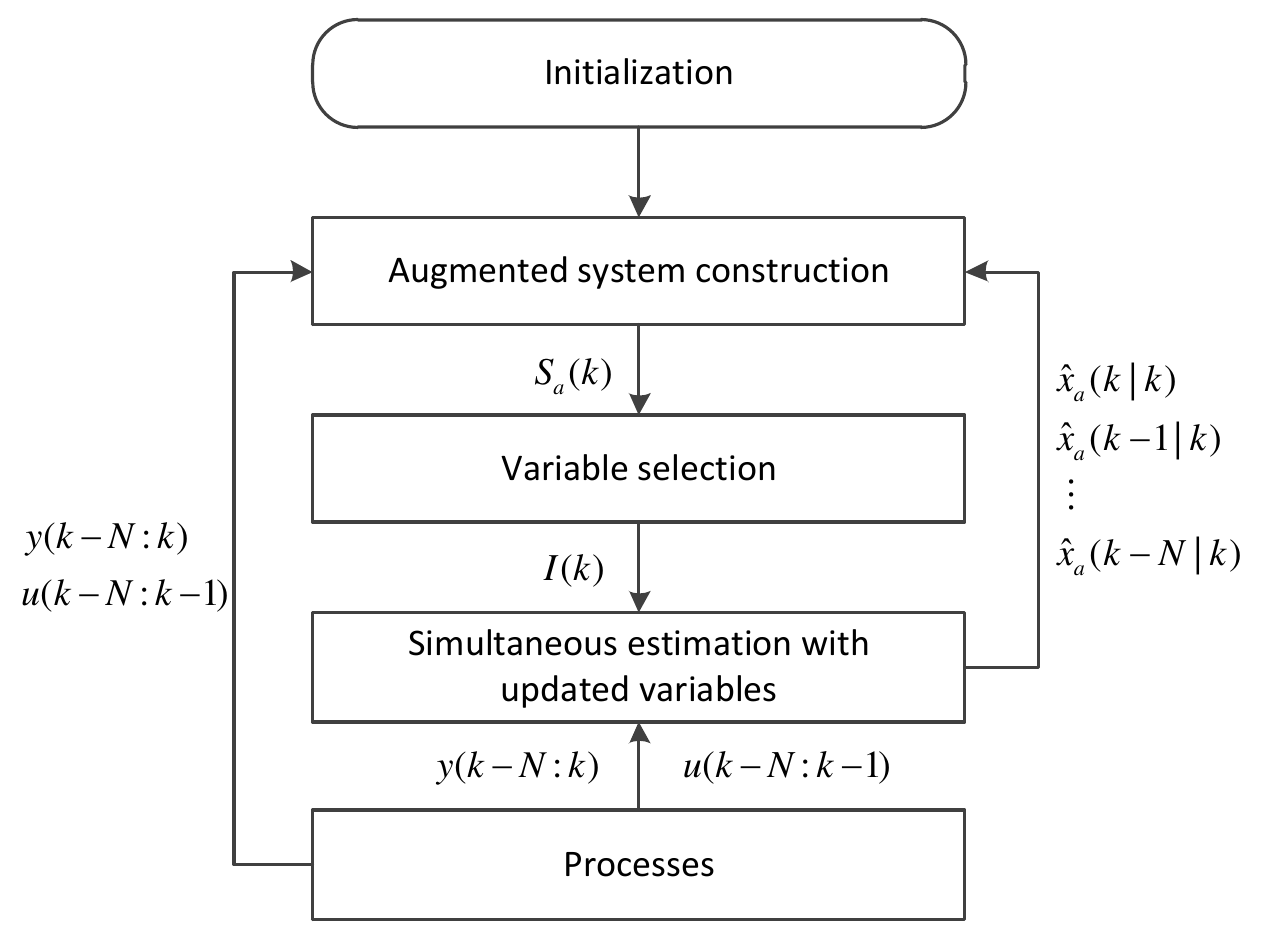}
  \caption{The implementation procedure and information flow of the proposed simultaneous state and parameter estimation integrated with sensitivity analysis.}\label{fig:6.0.1}
\end{figure}

In this section, we show how sensitivity analysis may be used to select an appropriate subset of states and parameters for estimation when not all the states and parameters can be estimated simultaneously. We will also discuss how sensitivity analysis may be integrated with simultaneous estimators to improve estimation performance. Figure~\ref{fig:6.0.1} shows the implementation procedure and information flow of the proposed approach for simultaneous state and parameter estimation integrated with sensitivity analysis. We will discuss the proposed approach in the framework of MHE. The proposed approach can be adopted to other estimation frameworks in a straightforward manner.

\subsection{Augmented system}\label{sec:6.1}

For simultaneous state and parameter estimation of system~(\ref{eq:2.1.1}), we consider augmenting the parameters as states, which is a rather standard approach in simultaneous state and parameter estimation \cite{plett2006sigma,bo2020parameter}. Following this, we can obtain the following augmented system:
\begin{subequations}\label{eq:6.0.1}
\begin{align}
    x_a(k+1) &= \left[\begin{array}{c} F(x(k),u(k),\theta(k))  \\  \theta(k) \end{array}\right]:=
F_a(x_a(k), u(k)) \label{eq:6.0.1a}\\
y(k) &= H(x(k), \theta(k)):=H_a(x_a(k)) \label{eq:6.0.1b}
\end{align}
\end{subequations}
where $x_a(k)=\left[ x(k)^T\; \theta(k)^T \right]^T \in R^{n+p}$ denotes the augmented state vector, $F_a(\cdot)$ and $H_a(\cdot)$ denote the augmented state and output equations, respectively.

The simultaneous state and parameter estimation objective is now to estimate the augmented state $x_a$ based on input and output information. First, we need to check whether the entire augmented state vector $x_a$ is observable. Based on the discussion in Section~\ref{sec:4}, we can check the rank of the following sensitivity matrix obtained following (\ref{eq:4.2.8}) along the typical trajectory within a data window of the augmented system:
\begin{equation}\label{eq:6.1.1}
    S_a(k) =  \left[ {\begin{array}{*{20}{l}}
    {{S_{y,x_a(k-N)}}(k-N)}\\
    {{S_{y,x_a(k-N)}}(k-N+1)}\\
    \kern 10pt \vdots \\
    {{S_{y,x_a(k-N)}}(k)}
    \end{array}} \right]
\end{equation}
where $N$ is the data window size and should be greater than or equal to $n+p$.


Based on the input and output data from processes, we can obtain the following expression for the evaluation of the sensitivity of the output $y(i), i=k-N, \cdots, k$ to the augmented state $x_a(k-N)$ following (\ref{eq:4.2.6}):
\begin{equation}\label{eq:6.1.3}
    S_{y,x_a(k-N)}(i) = {{\frac{{\partial H_a}}{{\partial x_a}}} (i)} {{\frac{{\partial F_a}}{{\partial x_a}}} (i-1)} {{\frac{{\partial F_a}}{{\partial x}}} (i-2)} \cdots {{\frac{{\partial F_a}}{{\partial x_a}}} (k-N)}
\end{equation}
where
\begin{equation*}
    {{\frac{{\partial F_a}}{{\partial x_a}}} (i)}=\left[\begin{array}{cc}\frac{\partial F}{\partial x}(i) & \frac{\partial F}{\partial \theta}\\ 0 & I_{p\times p}\end{array}\right], \quad \frac{\partial H_a}{\partial x_a}(i) = \left[\begin{array}{cc}\frac{\partial H}{\partial x}(i) & \frac{\partial H}{\partial \theta}\end{array}\right]
\end{equation*}

By checking the rank of the sensitivity matrix $S_a(k)$ at each sampling time, we can conclude whether the entire augmented state vector $x_a$ can be estimated locally using the input and output information. At the same time, we would like to note that by checking only the rank of the sensitivity matrix $S_a(k)$, it may not be sufficient especially when dealing with large-scale systems. The condition number of the matrix $S_a(k)$ should also be examined. Even when $S_a(k)$ is full rank, if its condition number is high, it may imply that the matrix is ill-conditioned and the states and parameters are difficult to be reliably estimated simultaneously. 

When the sensitivity matrix $S_a(k)$ is full rank along all the sampling points and is well conditioned, we may design an observer or estimator to estimate the states and parameters simultaneously. A more challenging case is that $S_a(k)$ is not full rank or is ill-conditioned. One approach to address this issue is, for example, to increase the number of measured output variables to make $S_a(k)$ full rank and well-conditioned. In this work, we assume that we do not have this option and focus on how we may select the most appropriate subset of the states and parameters for estimation.

\subsection{Variable selection}\label{sec:6.2}

Sensitivity analysis has been often used in parameter selection for parameter identification \cite{chu2012generalization,kravaris2013advances}. In this work, we adopt the idea of parameter (subset) selection into simultaneous state and parameter estimation, to select the most important state and parameter subset.

When $S_a(k)$ is not full rank along all the sampling points or is ill-conditioned, it is an indication that not all the elements in the augmented state vector $x_a$ can be estimated. In this case, one feasible approach is to only estimate those elements that are important in the prediction of the outputs. To select the most important elements, we can resort to the information contained in the sensitivity matrix $S_a(k)$. Specifically, we propose to use the orthogonalization method to select the most important elements for predicting $y$ from $x_a$. The objective is to find elements of $x_a$ that have little or no impact on the output $y$ (i.e., the sensitivities of $y$ to those elements of $x_a$ are very small). This is done by finding the strongly linearly independent columns in $S_a$ (each column corresponds to an element in $x_a$) and removing the columns that can be represented by those strongly independent columns or the columns that are weakly linearly independent on those strongly independent columns. Note that the sensitivity matrix $S_a(k)$ should be normalized with respect to the magnitudes of the different elements in $x_a$ before performing the orthogonalization method. Please refer to Remark \ref{remark:1} for possible approaches to normalize the sensitivity matrix $S_a(k)$. To find the strongly linearly independent columns, we may start with the column of the normalized $S_a(k)$ that has the biggest norm. Then, remove the information that can be expressed by the selected column and find the column that has the biggest norm in the remaining information (residual matrix). These steps can be repeated to find all the strongly linearly independent columns.

The detailed procedure to sequentially select the most important and estimable states and parameters for simultaneous estimation using the orthogonalization method, adapted from parameter selection for model identification \cite{yao2003modeling,lund2008parameter}, is shown below:
\begin{itemize}
    \item[S1:] At time instant $k$, evaluate the norm of each column of the normalized $S_a(k)$, initialize $j=1$ and select the column with the largest norm and denote it as $X_j$; 
    \item[S2:] Estimate the information in $S_a(k)$ that can be expressed by $X_j$: ${Z}_j = X_j ({X_j}^T X_j)^{-1} {X_j}^T S_a(k)$ and calculate the residual information/matrix: $R_j = S_a(k) - {Z}_j$. 
    \item[S3:] Evaluate the norm of each column of the residual matrix $R_j$; select the column from $S_a(k)$ that corresponds to the column with the largest norm in $R_{j}$; and add the selected column from $S_a(k)$ to $X_{j}$ as a new column to form $X_{j+1}$;
  \item[S4:] If the rank of $X_j$ is the same as the rank of $S_a(k)$ or the largest norm of the columns of $R_j$ is smaller than a prescribed cut-off value, then terminate the algorithm and the selected elements of $x_a$ correspond to the selected columns in $X_j$; otherwise, repeat S2-S4 with $j\leftarrow j+1$.
\end{itemize}





\begin{figure}
  \centering
  \includegraphics[width=0.45\textwidth,angle=0]{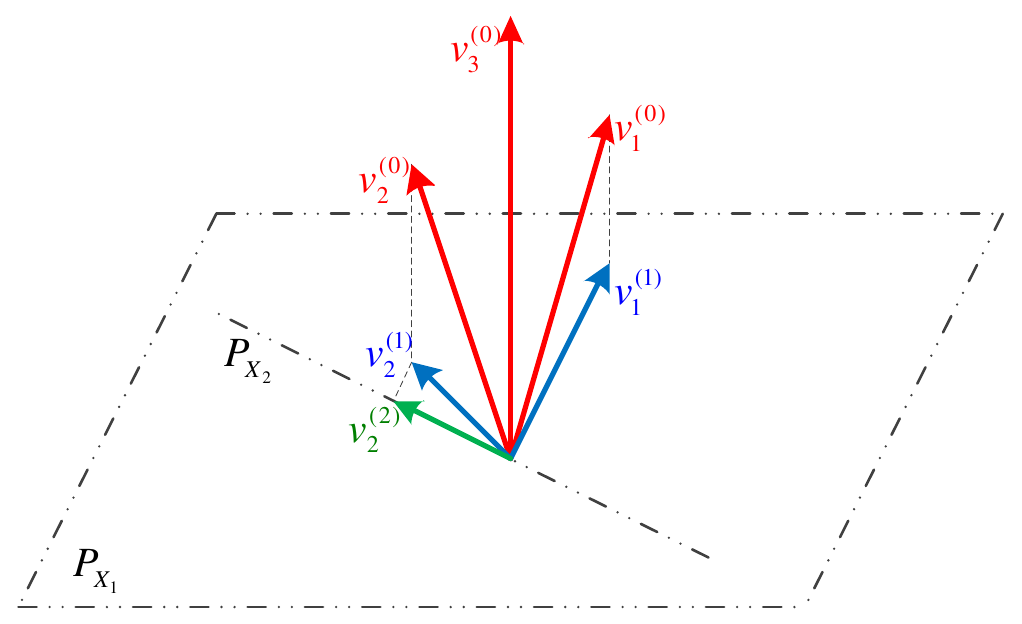}
  \caption{An illustration of the orthogonalization method.}\label{fig:6.2.1}
\end{figure}

Let us take an example to explain the above algorithm. Figure~\ref{fig:6.2.1} shows an illustration of this example. Suppose that there are in total three elements in $x_a$ and are denoted as $x_{a,i}$, $i = 1,\;2,\;3$, respectively. At time $k$, let us denote the corresponding column vectors to the three elements in $S_a(k)$ as $v_i^{(0)}$, $i = 1,\;2,\;3$, respectively. In the first step (S1), we evaluate the norm of each of the three column vectors. Suppose that $v_3^{(0)}$ has the largest norm. Then, $X_1=v_3^{(0)}$. In the second step (S2), we evaluate the information that cannot be expressed by $X_1$. To do this, we find the plane that is perpendicular to $X_1$ (denoted as $P_{X_1}$) and project the other vectors ($v_1^{(0)}$, $v_2^{(0)}$) to the $P_{X_1}$. These projected vectors on $P_{X_1}$, $v_1^{(1)}$, $v_2^{(1)}$, denote the information that cannot be represented by $X_1$ and this information can be expressed as $R_1=S_a(k)-Z_1$ with $Z_1=X_1(X_1^TX_1)^{-1}X_1^TS_a(k)$. In the third step (S3), we continue the analysis with $v_1^{(1)}$, $v_2^{(1)}$ (i.e., $R_1$). Within these vectors, we find the one with the largest norm. Suppose that the one is $v_1^{(1)}$. Then, we collect $v_1^{(0)}$ from $S_a(k)$ and add it to $X_1$ to form a new matrix with two columns $X_2=[v_3^{(0)}\; v_1^{(0)}]$. Subsequently, we evaluate the information that cannot be expressed by $X_2$. This can be done by projecting $v_2^{(1)}$ to the line/plane (denoted as $P_{X_2}$) that is perpendicular to $v_1^{(1)}$ within $P_{X_1}$. The projected vector on $P_{X_2}$, $v_2^{(2)}$, denotes the information that cannot be expressed by $X_2$ and this information can be expressed by the residual matrix $R_2$. Since we only have three elements in $x_a$, we now have a rank of the importance of elements in predicting $y$. That is, $x_{a,3}$ is the most important, and then $x_{a,1}$ and $x_{a,2}$ is the least important. If the rank of $S_a(k)$ is 2, we may only estimate two variables and $x_{a,3}$, $x_{a,1}$ should be the two variables that we estimate.



\subsection{MHE integrated with sensitivity analysis result}\label{sec:6.3}

At time instant $k$, after performing the variable selection based on the sensitivity matrix $S_a(k)$, we can determine the elements of $x_a$ that can be estimated based on the input and output information from $k-N$ to $k$. Let us use $I(k)$ to denote the set that contains the indices of the un-selected elements of $x_a$ using the variable selection algorithm presented in the previous subsection. For instance, in the previous example, $x_{a,1}$ and $x_{a,3}$ are selected as the variables to estimate, then $I(k)=\{2\}$. This information can be integrated into the MHE design to improve the estimation performance. In the proposed design, we consider that the estimation window used in MHE is the same as the data window size $N$ used in $S_a(k)$. The design of the proposed MHE at time $k$ is based on the augmented system (\ref{eq:6.0.1}) and is described as follows:
\begin{subequations}\label{eq:6.3.2}
  \begin{align}
      \min\limits_{\scriptstyle{{\hat x_a}(k - N), {\hat w_a}(\cdot)}} J & = \sum\limits_{i = k - N}^{k - 1} {\left\| {{\hat w_a}(i)} \right\|_{Q^{ - 1}}^2}  + \sum\limits_{j = k - N}^k {\left\| {\hat v(j)} \right\|_{{R^{ - 1}}}^2}   + V({\hat x_a}(k - N))\label{eq:6.3.2a}\\
      {\text{subject to:~}}&{\hat x_a}(i + 1) = {F_a}(\hat x_a(i),u(i)) + {\hat w_a}(i) \label{eq:6.3.2b}\\
    &{y(i) = H_a(\hat x_a(i)) + \hat v(i)} \label{eq:6.3.2c}\\
    &\hat x_a(i) \in X_a, \; \hat v(i)\in V, \; \forall i=k-N,\ldots,k \label{eq:6.3.2d}\\
    &\hat w_{a}(i) \in {W_{a}}, \; \forall i=k-N,\ldots,k-1 \label{eq:6.3.2e}\\
    &\hat x_{a,l}(k-N) = \hat x_{a,l}(k-N|k-1), \; l\in I(k) \label{eq:6.3.2f}\\
    &\hat w_{a,l}(i) = 0, \; l\in I(k),\; \forall i=k-N,\ldots,k-1 \label{eq:6.3.2g}
  \end{align}
\end{subequations}
where $\hat x_a$ denotes the estimated value of $x_a$, $\hat w_a$ denotes the estimated system disturbance, $\hat v$ denotes the estimated measurement noise, $X_a$, $W_a$ and $V$ denote the known constraints on the augmented state, the system disturbance, and the measurement noise. In (\ref{eq:6.3.2}), (\ref{eq:6.3.2a}) is the cost function the MHE tries to minimize, in which $Q^{-1}$, $R^{-1}$ are positive definite weighting matrices and $V(\hat x_a(k-N))$ is the arrival cost for the estimation problem. (\ref{eq:6.3.2b}) and (\ref{eq:6.3.2c}) are the system model with system disturbance and measurement noise considered. 
(\ref{eq:6.3.2d}) and (\ref{eq:6.3.2e}) are the known constraints on the state, measurement noise and system disturbance. (\ref{eq:6.3.2f}) and (\ref{eq:6.3.2g}) are the key constraints that take into account the variable selection results and they force the elements of the system disturbance vector corresponding to the unselected variables (not estimated variables) to be 0 and that the unselected variables evolute only according to the system model in an open-loop fashion with the initial condition $\hat x_{a,l}(k-N)$ specified as the value obtained (either estimated or predicted in open-loop) at the previous time instant. $I(k)$ is updated every time instant so (\ref{eq:6.3.2g}) should also be updated accordingly. Once the above optimization problem is solved, the optimal solution is denoted by $\hat x_a(l|k)$, $l=k-N,\ldots,k$. $\hat x_a(k|k)$ is the optimal estimate of $x_a$ for the current time $k$.

\begin{remark}\label{remark:1}
Note that before conducting variable selection, it is important to normalize the sensitivity matrix $S_a(k)$ with respect to the magnitudes of the different elements in $x_a$. One approach to normalize the sensitivity matrix is to multiply the elements in $S_a(k)$, $S_{y,x_a(k-N)}(l)$, $l=k-N,\ldots,k$, by $\frac{x_a(k-N)}{y(l)}$. Note also that when we evaluate the sensitivity in (\ref{eq:6.1.1}) following (\ref{eq:6.1.3}), values of $x_a$ along the window from $k-N$ to $k$ are needed. We can use the optimal estimates obtained at $k-1$, $\hat x_a(i|k-1)$, $i=k-N,\ldots,k-1$, and a prediction of $x_a(k)$ generated based on the augmented model: $x_a(k) = {F_a}(\hat x_a(k-1|k-1),u(k-1))$.
\end{remark}

\section{Application to a chemical process example}\label{sec:7}

In this section, we apply the proposed procedure to a benchmark chemical process to illustrate its applicability and effectiveness.

\subsection{Process description}\label{sec:7.1}

\begin{figure}
  \centering
  \includegraphics[width=0.45\textwidth,angle=0]{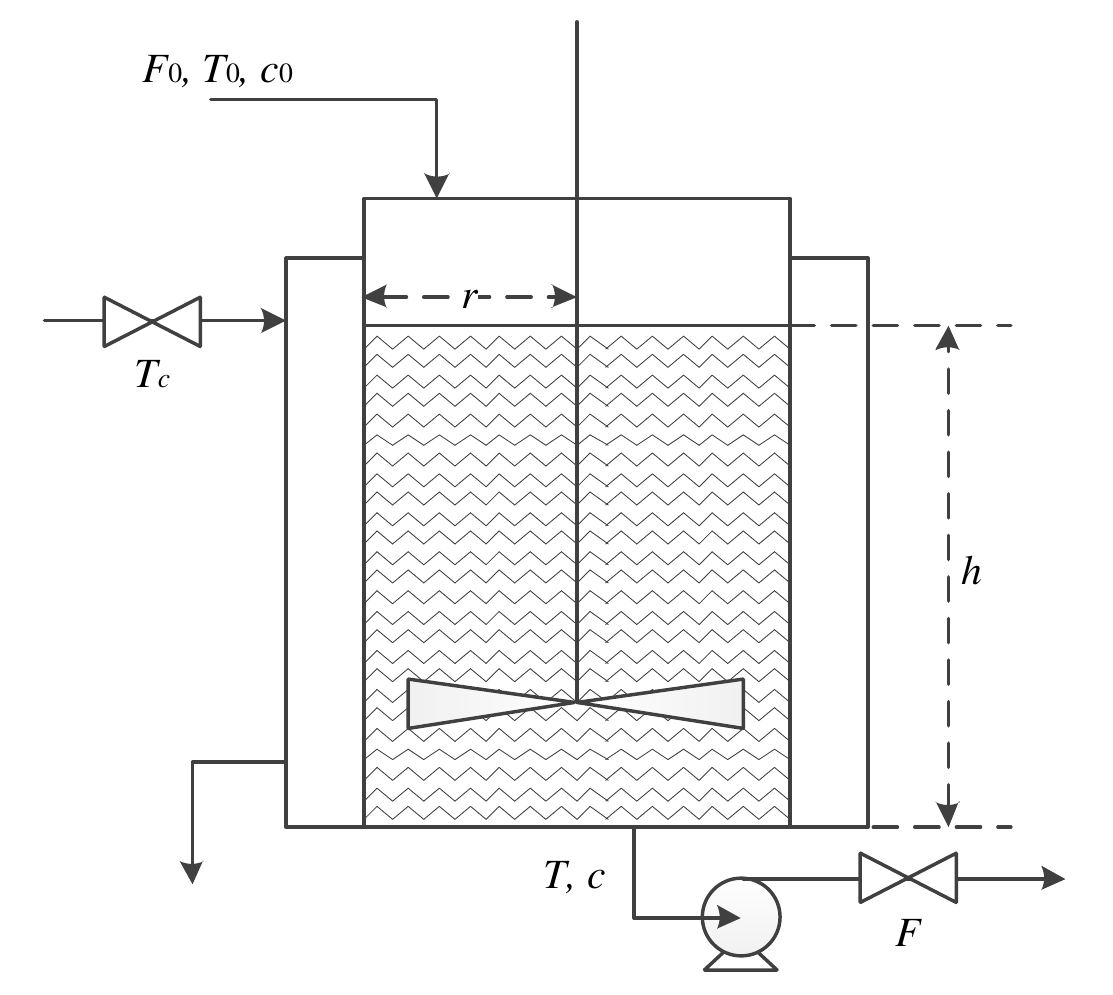}
  \caption{A schematic diagram of the CSTR process.}\label{fig:7.1.1}
\end{figure}

We consider a nonlinear continuous stirred-tank reactor (CSTR) as shown in Figure~\ref{fig:7.1.1}. In the CSTR, an irreversible first-order reaction $A \rightarrow B$ takes place. The CSTR is equipped with an external cooling jacket for temperature regulation purpose. The dynamics of the CSTR are described as follows \cite{pannocchia2003disturbance,rawlings2009model}:
\begin{subequations}\label{eq:7.1.1}
\begin{align}
\frac{{dc}}{{dt}} &= \frac{{{F_0}({c_0} - c)}}{{\pi {r^2}h}} - {k_0}\exp ( - \frac{E}{{RT}})c \label{eq:7.1.1a}\\
\frac{{dT}}{{dt}} &= \frac{{{F_0}({T_0} - T)}}{{\pi {r^2}h}} + \frac{{ - \Delta H}}{{\rho {C_p}}}{k_0}\exp ( - \frac{E}{{RT}})c + \frac{{2U}}{{r\rho {C_p}}}({T_c} - T) \label{eq:7.1.1b}\\
\frac{{dh}}{{dt}} &= \frac{{{F_0} - F}}{{\pi {r^2}}} \label{eq:7.1.1c}
\end{align}
\end{subequations}
In the above model, $c$ is the molar concentration of the reactant $A$; $T$ is the reactor temperature; $h$ is the liquid level in the reactor; $T_c$ is the coolant temperature; $F$ is the outlet flow rate of the CSTR;  $F_0$, $T_0$, and $c_0$ denote the flow rate, temperature, and molar concentration of the feed to the CSTR, respectively; $r$ denotes the radius of the reactor floor; $\Delta H$, $k_0$, and $E$ denote the enthalpy, pre-exponential constant, and activation energy of the reaction, respectively; $R$ denotes the gas constant; $U$ denotes the heat transfer coefficient; $C_p$ and $\rho$ denote the heat capacity and density of the fluid in the reactor, respectively. The values of the process parameters are shown in Table~\ref{tab:7.1.1}. The continuous model is discretized using the fourth order Runge-Kutta method with a sampling time $\Delta T = 0.2$ min. At each sampling point, $h$ and $T$ are measured.

\begin{table}
  \caption{Values of the parameters of the CSTR.}\label{tab:7.1.1}
  \begin{center}
    \begin{tabular}{lll}
    \hline
      Paramter & Nominal value &      Units \\
    \hline
    $F_0$           &        0.1 &           $\rm{m^3/min}$ \\

    $T_0$           &        350 &           $\rm{K}$ \\

    $c_0$           &          1 &           $\rm{kmol/m^3}$ \\

    $r$           &      0.219 &           $\rm{m}$ \\

    $k_0$           &  $7.2 \times 10^{10}$ &           $\rm{min^{-1}}$ \\

    $E/R$           &       8750 &           $\rm{K}$ \\

    $U$           &      54.94 &           $\rm{kJ/min} \cdot \rm{m^2} \cdot \rm{K}$ \\

    $\rho$           &       1000 &           $\rm{kg/m^3}$ \\

    $C_p$           &      0.239 &           $\rm{kJ/kg} \cdot \rm{K}$ \\

    $\Delta H$           &  $-5 \times 10^4$ &           $\rm{kJ/kmol}$ \\
    \hline
    \end{tabular}
\end{center}
\end{table}

It is assumed that these parameters of the process are not known exactly in the design of the MHE, and we want to estimate (some or all of) the state variables ($x=[c,\;T,\;h]^T$) and some of the parameters of the process based on the two output measurements. The objective is to extract as much information as possible from the two output measurements and get the best possible estimation performance. We will illustrate how the proposed variable selection and estimation methods may be used to achieve this objective.

\subsection{Augmented system construction and simulation settings}\label{sec:7.2}

The first step is to construct the augmented system. Only some of the parameters that are uncertain are considered in the augmented system. Specially, the parameter set considered is
$\theta=[F_0,T_0,c_0,k_0,E/R,U,C_p,\Delta H]^T$. The augmented state $x_a$ is as follows:
\begin{equation}\label{eq:7.2.1}
    x_a = [c,T,h,F_0,T_0,c_0,k_0,E/R,U,C_p,\Delta H]^T
\end{equation}
which contains the three original states and eight parameters. As described above, the output is $y=[T, h]^T$. Corresponding to the parameters, the process has a steady-state:
\[x_s=[0.878~{\rm {kmol/m^3}}, 324.5~{\rm K}, 0.659~{\rm m}]^T\]
To avoid the potential influence of parameter tuning and numerical tolerance in rank calculation, let us consider that the augmented model is normalized around the steady-state and the parameter values with respect to the following element-wise state constraints on the augmented state:
\begin{equation*}
  x_{a,s}-0.3 |x_{a,s}| \leq \hat x_a(k) \leq x_{a,s}+0.3 |x_{a,s}|
\end{equation*}
where $x_{a,s}$ denotes the steady-state augmented with the parameter values.

\begin{figure}
    \centerline{\includegraphics[width=0.6\textwidth]{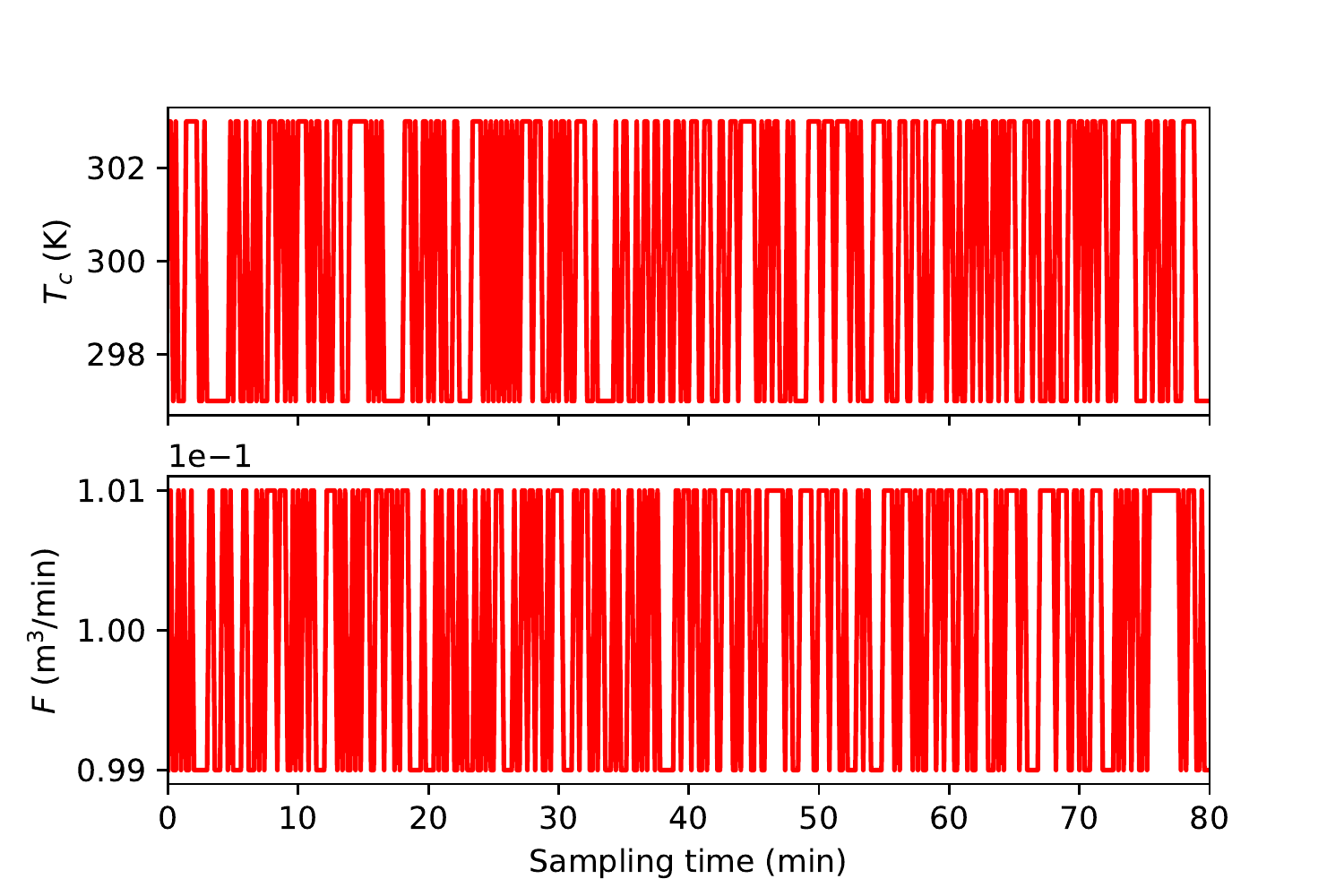}}
  \caption{Trajectories of the two manipulated inputs in the simulations.}\label{fig:7.2.1}
\end{figure}

In the following simulations, the actual process data is generated with the outlet flow rate $F$ and the coolant temperature $T_c$ taking random binary sequences as shown in Figure~\ref{fig:7.2.1}. The initial values of the three states are their steady-state values. Gaussian process noise $w$ with zero mean and standard deviation $0.6 \times 10^{-3} |x_{a,s}|$ are added on the three original states and Gaussian measurement noise $v$ with zero mean and standard deviation $0.6 \times 10^{-3} |y_{a,s}|$ are added on the measurements.

In the variable selection algorithm, a pre-determined cut-off value is needed for the termination of the algorithm. We propose to use the following cut-off value:
\begin{equation}\label{eq:7.3.1}
  \lambda = \alpha \sqrt{\sigma_w^2 + \sigma_v^2}
\end{equation}
where $\alpha$ is a tuning coefficient, $\sigma_w^2$ and $\sigma_v^2$ are the variances of the process noise and measurement noise, respectively. The design of this cut-off value is to use the summation of the process and measurement noise variances to approximate for the noise vairance in the sensitivity matrix.

In the design of the MHE, to avoid the potential bias caused by the arrival cost design in the estimation performance, the estimation window $N$ is chosen to be the same as the length of the total simulation time. This implies that the MHE uses all the available measurements from initial time 0 and is equivalent to the full information estimation (FIE). In the MHE, the values of $Q$ and $R$ are the same as the variances of the process and measurement noise.

To assess the estimation performance, a few indexes are used. One performance index is the average relative standard deviation ${\sigma}_{x_{a,i}}, i = 1,2,\cdots,11$:
\begin{equation}
    {\sigma}_{x_{a,i}} = \sqrt {\frac{{\sum\nolimits_{k = 0}^{{N_{sim}} - 1} {{{\left( {({{\hat x}_{a,i}}(k) - {x_{a,i}}(k))/{x_{a,i}}(k)} \right)}^2}} }}{{{N_{sim}}}}}  \label{eq:7.3.2a}
\end{equation}
where $N_{sim}$ indicates the total simulation time/steps, $\hat x_{a,i}$ denotes the estimated value and $x_{a,i}$ denotes the actual value of the $i$-th element in the augmented state. A couple other performance indexes are the root mean square error (RMSE) at a time instant and the everage RMSE:
\begin{subequations}\label{eq:7.3.2}
  \begin{align}
    {\rm{RMSE}}_{x_a}(k) &= \sqrt {\frac{{\sum\nolimits_{i = 1}^{n_{x_a}} {{{\left( {({{\hat x}_{a,i}}(k) - {x_{a,i}}(k))/{x_{a,i}}(k)} \right)}^2}} }}{{{n_{x_a}}}}} \label{eq:7.3.2b} \\
    {{\rm{RMSE}}_{x_a}} &= \frac{\sum\nolimits_{k = 0}^{N_{sim}-1}{\rm{RMSE}}_{x_a}(k)}{N_{sim}}  \label{eq:7.3.2c}
  \end{align}
\end{subequations}
where ${\rm{RMSE}}_{x_a}(k)$ with $k=0, \ldots, N_{sim}-1$ shows the evolution of the RMSE value over time and ${\rm{RMSE}}_{x_a}$ shows the average value.

\subsection{Results}\label{sec:7.3}

First, to verify the effectiveness of the proposed method, we conduct the following experiments. Specifically, three different cases are considered. In Case 1, all the 11 variables in the augmented state are estimated simultaneously; that is, $I(k)=\emptyset$, for all $k$. In Case 2, the proposed variable selection algorithm is used to select the most important and estimable variables based on sensitivity analysis and $I(k)$ is obtained according to the algorithm. In Case 3, we consider that the three original states are important and must be estimated at each sampling time and variable selection is only performed among the parameters. In Case 3, at each sampling time we remove the information that can be expressed by the three original states from the obtained sensitivity matrix $S_a(k)$ first and then use the residual matrix to sequentially select a few parameters to estimate simultaneously with the three states following a similar procedure as shown in Section \ref{sec:6.2}.

\begin{figure}
  \centering
    \subfigure[]{
    \label{fig:7.3.1a} 
    \includegraphics[width=0.48\textwidth,angle=0]{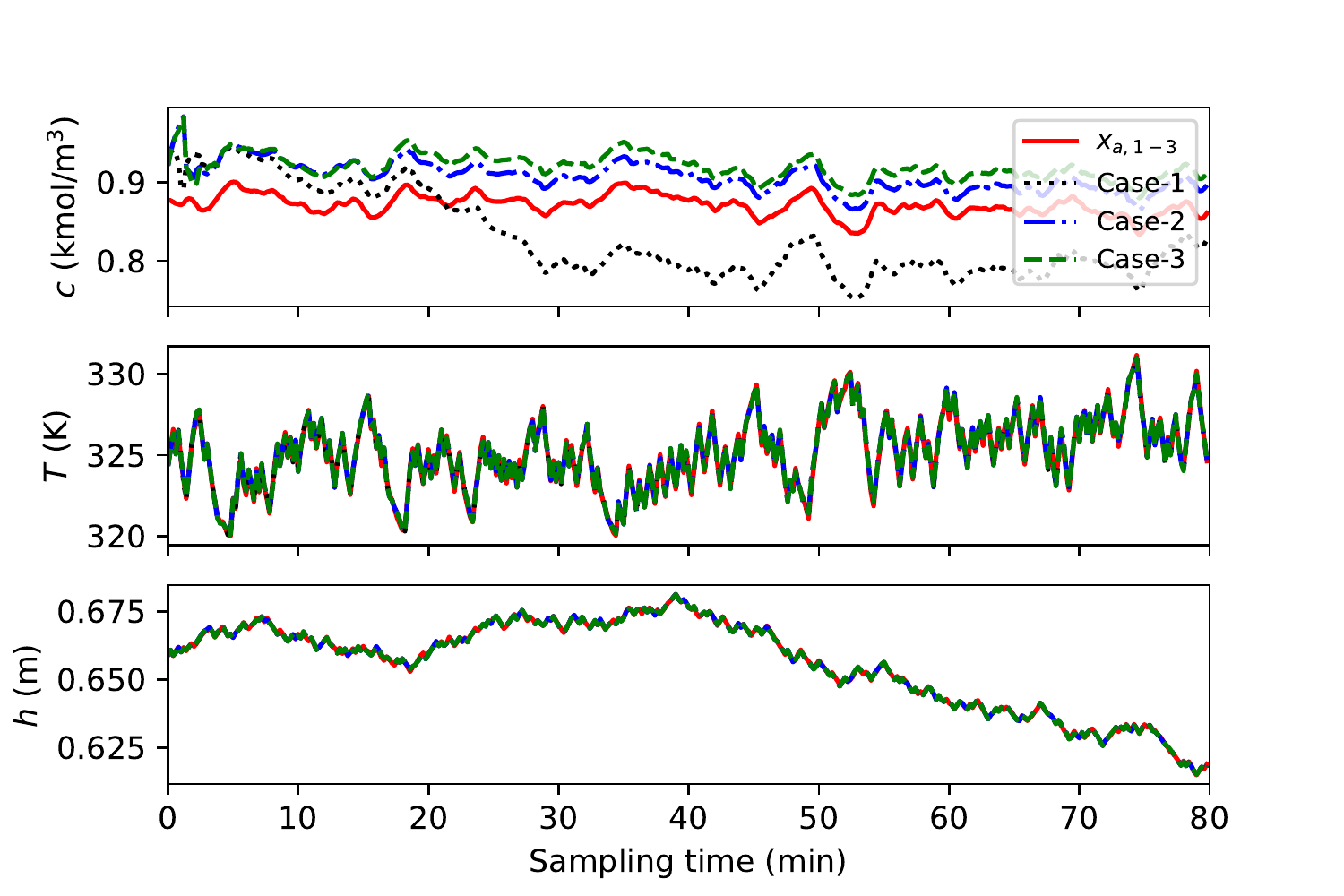}}
    \hspace{0in}
    \subfigure[]{
    \label{fig:7.3.1b} 
    \includegraphics[width=0.48\textwidth,angle=0]{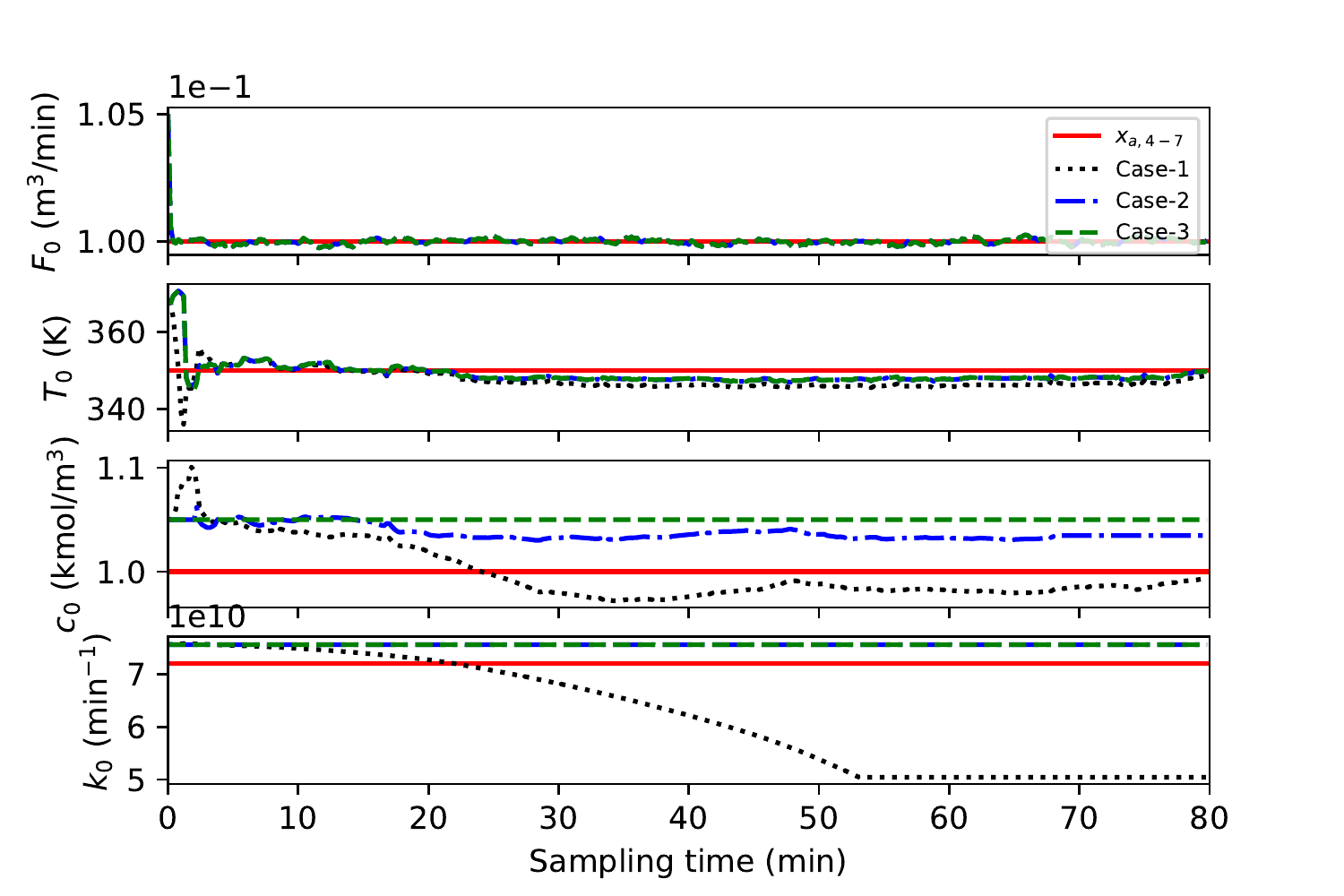}}
    \hspace{0in}
    \subfigure[]{
    \label{fig:7.3.1c} 
    \includegraphics[width=0.48\textwidth,angle=0]{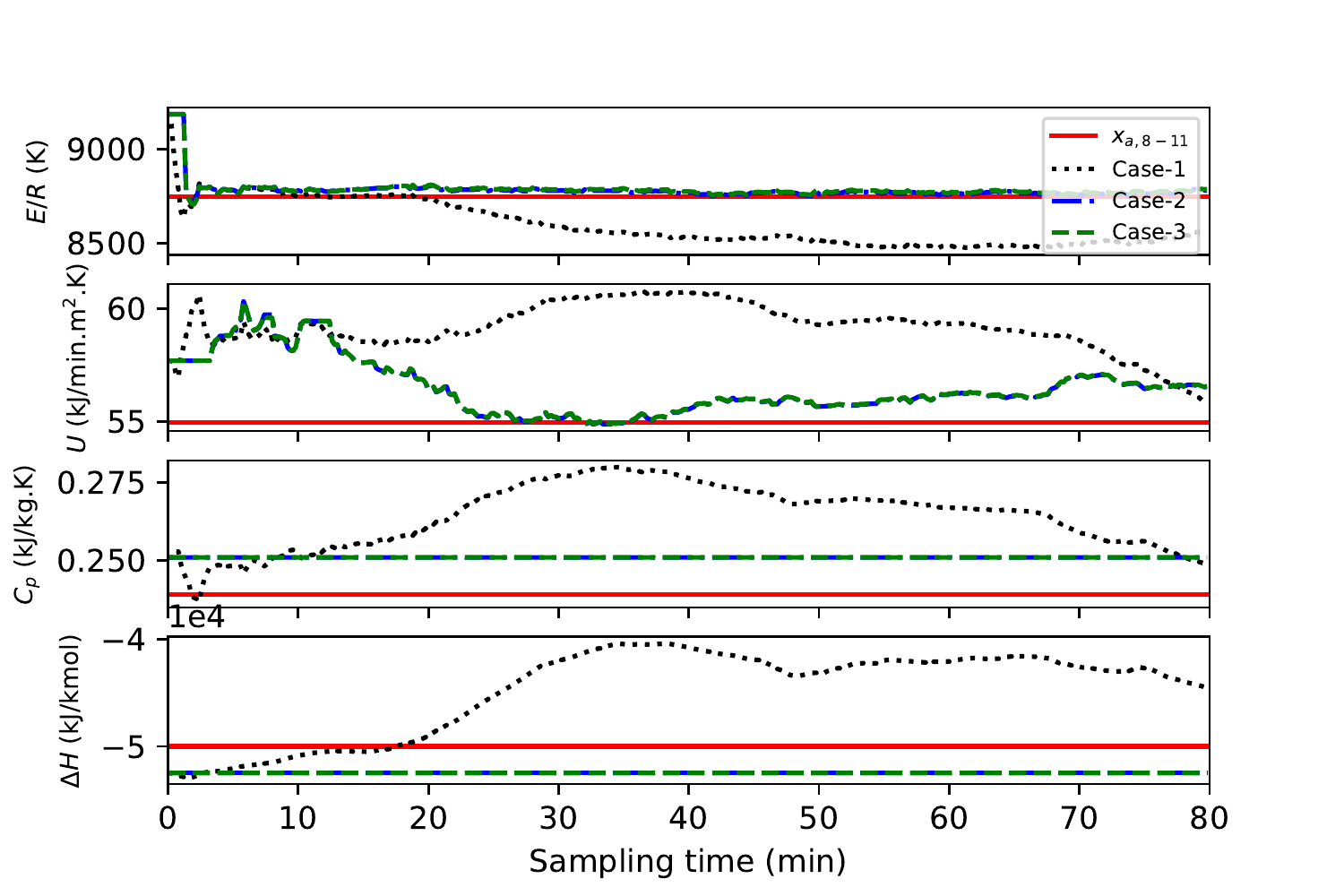}}
    \hspace{0in}
    \subfigure[]{
    \label{fig:7.3.1d} 
    \includegraphics[width=0.48\textwidth,angle=0]{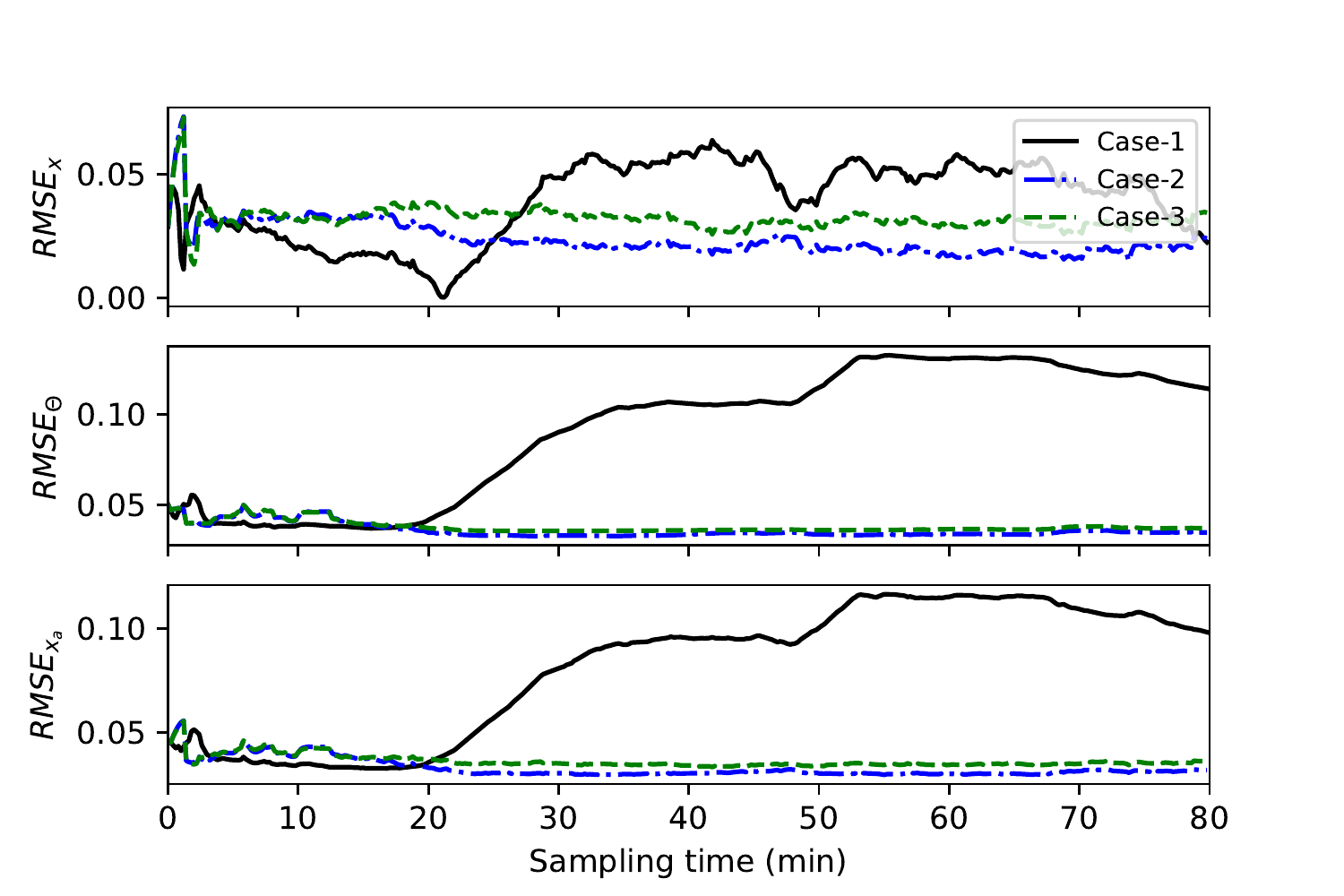}}
    \caption{(a)-(c) Trajectories of the actual states and parameters (solid lines), estimated states and parameters in Case 1 (dotted lines), estimated states and parameters in Case 2 (dash dotted lines), and estimated states/parameters in Case 3 (dashed lines). (d) Evolution of the RMSE of the three original state vector, the parameter vector, and the entire augmented state vector during the simulation time in Case 1 (solid lines), Case 2 (dash dotted lines) and Case 3 (dashed lines).}\label{fig:7.3.1}
\end{figure}

\begin{figure}
    \centerline{\includegraphics[width=0.6\textwidth,angle=0]{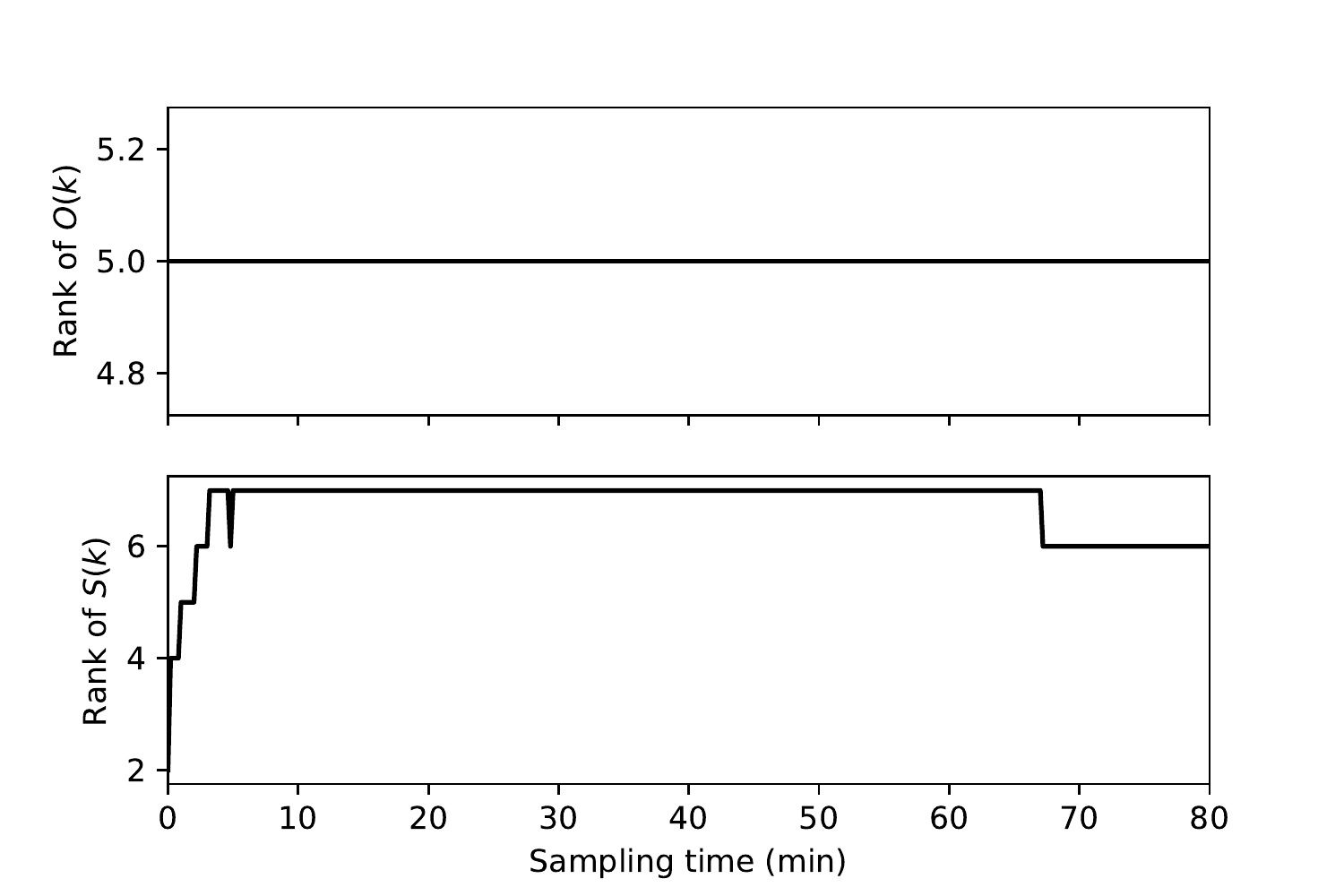}}
    \caption{Rank of the observability matrix (\ref{eq:4.2.4}) based on the linearized models of the augmented system along the actual trajectory (top) and the rank of the sensitivity matrix (\ref{eq:4.2.8}) of the augmented system along the actual trajectory (bottom).}\label{fig:7.3.1e}
\end{figure}

In this set of simulations, we use $\alpha = 2$. In the MHE design, a 5\% mismatch in the initial state of each of the three original states is considered. It is also assumed that the parameters are not known exactly and there is a 5\% mismatch in each of the parameters. The simulation results of the three cases are shown in Figure~\ref{fig:7.3.1}. From Figure~\ref{fig:7.3.1}, it can be seen that the estimation performance in Case 1 is much poorer compared with Case 2 and Case 3. The poor estimation performance of Case 1 is due to the unobservability of the entire augmented state vector. The rank of the observability matrix (\ref{eq:4.2.4}) based on the linearized models of the augmented system and the rank of the sensitivity matrix (\ref{eq:4.2.8}) of the augmented system along the actual trajectory of the system are shown in Figure~\ref{fig:7.3.1e}. From the figure, it can be seen that the rank of either matrix is much smaller than 11 at any time instant. This implies that the full augmented state $x_a$ is not observable. Estimating $x_a$ without considering the observability may lead to overfitting the outputs and poor state estimation performance as shown in Case 1.

In Case 2 and Case 3, the observability information is taken into account and only a subset of the variables selected based on the sensitivity matrix is estimated. In Case 2, in the selection of the subset of variables, the original states and the parameters are treated equally while in Case 3, preference is given to the original states and they are included in the subset all the time for estimation and only the parameters go through the selection process as described earlier. From Figure~\ref{fig:7.3.1}, it can be seen that the estimation results of Case 2 and Case 3 are similar. This can be further seen from Table~\ref{tab:7.3.1}, which summarizes the simulation results and shows the average performance indexes for the entire simulation. From Table~\ref{tab:7.3.1}, it can be seen that the performance of Case 2 is slightly better than Case 3. The performance difference can be explained by looking into the variables estimated in Case 2 and Case 3. Table~\ref{tab:7.3.2} shows the number of sampling times that a variable is included in the corresponding MHE estimation problem. From Table~\ref{tab:7.3.2}, it can be seen that in Case 2, one of the original state $x_{a,1}$ (i.e., $c$) is not included for estimation but one more parameter $x_{a,6}$ (i.e., $c_0$) is included in estimation for some time instants. The selected variables are strictly according to their sensitivity information. However, in Case 3, $x_{a,1}$, which is difficult to estimate based on the outputs, is always included for estimation and $x_{a,6}$ is never included for estimation. The results of Case 2 and Case 3 implies that the proposed variable selection according to the sensitivity information does lead to improved estimation performance.

\begin{table}
    \caption{The average performance indexes for the entire simulation of the three cases.}\label{tab:7.3.1}
  \centering
    \begin{tabular}{rrrrrrrr}
    \hline
               &        $\sigma_{x_{a,1}}$ &        $\sigma_{x_{a,2}}$ &        $\sigma_{x_{a,3}}$ &        $\sigma_{x_{a,4}}$ &        $\sigma_{x_{a,5}}$ &        $\sigma_{x_{a,6}}$ &        $\sigma_{x_{a,7}}$ \\
    \hline
        Case-1 &       7.40\% &       0.05\% &       0.05\% &       0.27\% &       1.04\% &       2.73\% &      19.83\% \\

        Case-2 &       4.29\% &       0.05\% &       0.05\% &       0.27\% &       0.92\% &       3.79\% &       5.00\% \\

        Case-3 &       5.62\% &       0.05\% &       0.05\% &       0.27\% &       0.91\% &       5.00\% &       5.00\% \\

    \hline
               &        $\sigma_{x_{a,8}}$ &        $\sigma_{x_{a,9}}$ &       $\sigma_{x_{a,10}}$ &       $\sigma_{x_{a,11}}$ &       RMSE$_{x}$ &       RMSE$_\theta$ &       RMSE$_{x_a}$ \\
    \hline
        Case-1 &       2.21\% &       7.93\% &      11.33\% &      13.25\% &       3.97\% &       9.15\% &       8.09\% \\

        Case-2 &       0.73\% &       3.59\% &       5.00\% &       5.00\% &       2.38\% &       3.58\% &       3.30\% \\

        Case-3 &       0.75\% &       3.61\% &       5.00\% &       5.00\% &       3.21\% &       3.77\% &       3.63\% \\

    \hline
    \end{tabular}
\end{table}

\begin{table}
    \caption{The number of sampling times that each variable is included in MHE in the three cases.}\label{tab:7.3.2}
  \centering
    \begin{tabular}{rrrrrrrrrrrr}
    \hline
               &          $x_{a,1}$ &          $x_{a,2}$ &          $x_{a,3}$ &          $x_{a,4}$ &          $x_{a,5}$ &          $x_{a,6}$ &          $x_{a,7}$ &          $x_{a,8}$ &         $x_{a,9}$ &          $x_{a,10}$ &          $x_{a,11}$ \\
    \hline
        Case-1 &        400 &        400 &        400 &        400 &        400 &        400 &        400 &        400 &        400 &        400 &        400 \\

        Case-2 &          0 &        400 &        400 &        399 &        399 &        331 &          0 &        393 &        367 &          0 &          0 \\

        Case-3 &        399 &        400 &        400 &        399 &        398 &          0 &          0 &        393 &        369 &          0 &          0 \\
    \hline
    \end{tabular}
\end{table}

Next, we perform another set of simulations to investigate the maximum number of variables that can be estimated simultaneously for the considered process. In this set of simulations, instead of using a cut-off value in the proposed variable selection algorithm, we include the first $n=4,5,6,7,8$ selected variables respectively in the MHE estimation all the time. The number of the estimated variables is not adjusted based on the sensitivity. The simulation results are summarized in Table~\ref{tab:7.4.1}. From Table \ref{tab:7.4.1}, it can be seen that the `best' estimation performance is obtained when $n=7$. Indeed, when $n=5,6,7$, the estimation performance is relatively close. When $n=4$, the estimation performance is obviously poorer. This is because that the number of variables is not sufficient to extract/represent the information contained in the two outputs. Similarly, when $n=8$, the performance is also obviously poorer. This is because that too many variables are included in the estimation and overfitting occurs. If we compare the above results with the results of Case 2 in the previous set of simualtions, we can see that the proposed approach indeed leads to a better estimation performance. This may be because that in the proposed approach the number of the estimated variables is not fixed and is determined based on the actual sensitivity information at each time instant.

\begin{table}
  \caption{The average performance indexes for the entire estimation with different $n$ values}\label{tab:7.4.1}
  \centering
    \begin{tabular}{rrrrrrrr}
    \hline
               &        $\sigma_{x_{a,1}}$ &        $\sigma_{x_{a,2}}$ &        $\sigma_{x_{a,3}}$ &        $\sigma_{x_{a,4}}$ &        $\sigma_{x_{a,5}}$ &        $\sigma_{x_{a,6}}$ &        $\sigma_{x_{a,7}}$ \\
    \hline
           $n=4$ &      15.77\% &       0.07\% &       0.05\% &       0.27\% &       5.33\% &       5.00\% &       5.00\% \\

           $n=5$ &       5.39\% &       0.05\% &       0.05\% &       0.27\% &       0.54\% &       5.00\% &       5.00\% \\

           $n=6$ &       6.76\% &       0.05\% &       0.05\% &       0.27\% &       0.66\% &       6.25\% &       5.00\% \\

           $n=7$ &       4.33\% &       0.05\% &       0.05\% &       0.27\% &       0.76\% &       4.25\% &       5.00\% \\

           $n=8$ &       5.07\% &       0.05\% &       0.05\% &       0.27\% &       0.94\% &       9.75\% &       5.00\% \\
    \hline
               &        $\sigma_{x_{a,8}}$ &        $\sigma_{x_{a,9}}$ &       $\sigma_{x_{a,10}}$ &       $\sigma_{x_{a,11}}$ &       RMSE$_{x}$ &       RMSE$_\theta$ &       RMSE$_{x_a}$ \\
    \hline
           $n=4$ &       5.00\% &       5.00\% &       5.00\% &       5.00\% &       9.06\% &       4.72\% &       6.22\% \\

           $n=5$ &       0.52\% &       5.00\% &       5.00\% &       5.00\% &       3.10\% &       3.96\% &       3.75\% \\

           $n=6$ &       0.54\% &       6.48\% &       5.00\% &       5.00\% &       3.88\% &       4.42\% &       4.29\% \\

           $n=7$ &       0.48\% &       5.01\% &       6.59\% &       5.00\% &       2.41\% &       4.12\% &       3.74\% \\

           $n=8$ &       1.12\% &       4.84\% &       6.74\% &      22.05\% &       2.87\% &       8.67\% &       7.59\% \\
    \hline
    \end{tabular}
\end{table}

Further, we carry out a set of simulations to study the impact of the tuning coefficient $\alpha$ in the cut-off value expressed in (\ref{eq:7.3.1}). We consider that $\alpha=1,2,3,4,5$, respectively. The simulation results are summarized in Table~\ref{tab:7.4.2}. From Table \ref{tab:7.4.2}, we can see that the proposed approach works well with different $\alpha$ values and the performance variation is minor. The best performance is achieved when $\alpha =2$. The proposed method could give us a clear and reliable guidance on which group variables we should estimate at each sampling time.

\begin{table}
  \caption{The average performance indexes for the entire estimation with different $\alpha$ values}\label{tab:7.4.2}
  \centering
    \begin{tabular}{rrrrrrrr}
    \hline
               &        $\sigma_{x_{a,1}}$ &        $\sigma_{x_{a,2}}$ &        $\sigma_{x_{a,3}}$ &        $\sigma_{x_{a,4}}$ &        $\sigma_{x_{a,5}}$ &        $\sigma_{x_{a,6}}$ &        $\sigma_{x_{a,7}}$ \\
    \hline
           $\alpha = 1$ &       4.29\% &       0.05\% &       0.05\% &       0.27\% &       0.75\% &       4.15\% &       5.00\% \\

           $\alpha = 2$ &       4.29\% &       0.05\% &       0.05\% &       0.27\% &       0.92\% &       3.79\% &       5.00\% \\

           $\alpha = 3$ &       4.98\% &       0.05\% &       0.05\% &       0.27\% &       0.78\% &       4.54\% &       5.00\% \\

           $\alpha = 4$ &       4.93\% &       0.05\% &       0.05\% &       0.27\% &       0.82\% &       4.45\% &       5.00\% \\

           $\alpha = 5$ &       5.65\% &       0.05\% &       0.05\% &       0.27\% &       0.93\% &       5.00\% &       5.00\% \\
    \hline
               &        $\sigma_{x_{a,8}}$ &        $\sigma_{x_{a,9}}$ &       $\sigma_{x_{a,10}}$ &       $\sigma_{x_{a,11}}$ &       RMSE$_{x}$ &       RMSE$_\theta$ &       RMSE$_{x_a}$ \\
    \hline
           $\alpha = 1$ &       0.48\% &       4.89\% &       6.46\% &       5.00\% &       2.40\% &       4.06\% &       3.69\% \\

           $\alpha = 2$ &       0.73\% &       3.59\% &       5.00\% &       5.00\% &       2.38\% &       3.58\% &       3.30\% \\

           $\alpha = 3$ &       0.73\% &       5.00\% &       5.00\% &       5.00\% &       2.83\% &       3.90\% &       3.64\% \\

           $\alpha = 4$ &       0.77\% &       5.00\% &       5.00\% &       5.00\% &       2.79\% &       3.89\% &       3.63\% \\

           $\alpha = 5$ &       0.89\% &       5.00\% &       5.00\% &       5.00\% &       3.20\% &       3.98\% &       3.79\% \\
    \hline
    \end{tabular}
\end{table}

\section{Conclusions}\label{sec:8}

In the work, the role of sensitivity analysis in simultaneous state and parameter estimation was discussed in detail. It was demonstrated that sensitivity analysis provides a way to check the observability of nonlinear systems and can be used to select variables for simultaneous estimation. This is especially useful and important for cases that the entire augmented system is not fully observable. In this work, an approach to integrate the results of variable selection into the framework of MHE was proposed. The results of extensive simulations demonstrated the performance of the proposed approach.

\section{Acknowledgement}

The first author, Jianbang Liu, was a visiting PhD student in the Department of Chemical and Materials Engineering at the University of Alberta from September 2018 to August 2020. He acknowledges the financial support from the China Scholarship Council (CSC) during this period. The author Tao Zou acknowledges the financial support from the Key-Area Research and Development Program of Guangdong Province under the grant number 2020B0101050001.


\end{document}